\documentclass[aps,prl,twocolumn,showpacs,superscriptaddress,nofootinbib]{revtex4-1}  
\usepackage{graphicx,color}  
\usepackage{bm}        
\usepackage{amssymb} 
\usepackage{amsmath}

\usepackage{graphics}
\usepackage{epsfig}
\usepackage{slashed}
\usepackage[utf8]{inputenc}
\usepackage{multirow}
\usepackage{pstricks}
\usepackage{dcolumn}
\usepackage{epsfig}
\usepackage{array}
\usepackage{booktabs}
\usepackage{slashed}

\def\lsim{\raise0.3ex\hbox{$\;<$\kern-0.75em\raise-1.1ex\hbox{$\sim\;$}}}
\def\gsim{\raise0.3ex\hbox{$\;>$\kern-0.75em\raise-1.1ex\hbox{$\sim\;$}}}

\newcommand{\be}{\begin{eqnarray}}
\newcommand{\ee}{\end{eqnarray}}
\def\bea{\begin{eqnarray}}
\def\eea{\end{eqnarray}}
\def\nn{\nonumber}
\def\Be{{}^8\textrm{Be}}

\begin{document}
\title{\fontsize{10.9}{10.9}\selectfont{Explanation of the 17 MeV Atomki Anomaly
 in a $U(1)^\prime $-Extended 2-Higgs Doublet Model } }
 
\author{Luigi Delle Rose} \affiliation{School of Physics and Astronomy, University of Southampton, Highfield, Southampton SO17 1BJ, UK} \affiliation{Particle Physics Department, Rutherford Appleton Laboratory, Chilton, Didcot, Oxon OX11 0QX, United Kingdom}
\author{Shaaban Khalil} \affiliation{Center for Fundamental Physics, Zewail City of Science and Technology, 6 October City, Giza 12588, Egypt}
\author{Stefano Moretti} \affiliation{School of Physics and Astronomy, University of Southampton, Highfield, Southampton SO17 1BJ, UK} \affiliation{Particle Physics Department, Rutherford Appleton Laboratory, Chilton, Didcot, Oxon OX11 0QX, United Kingdom}

\date{\today}

\begin{abstract}
Motivated by an anomaly observed in the decay of an excited state of Beryllium ($\Be$) by the Atomki collaboration, we study an
extension of the Standard Model with a gauged $U(1)'$ symmetry in presence of a 2-Higgs Doublet Model structure of the Higgs sector. We show that this scenario complies with a variety of experimental results and is able to explain the potential presence of a resonant spin-1 gauge boson, $Z'$, with a mass of 17 MeV in the Atomki experimental data, for appropriate choices of $U(1)'$ charges and Yukawa interactions.  
\end{abstract}
\maketitle

The Atomki pair spectrometer experiment \cite{Krasznahorkay:2015iga} was 
set up for searching $e^+e^-$ internal pair creation in the decay of excited $^8{\rm Be}$ nuclei (henceforth, $^8{{\rm Be}^*}$), the latter being
 produced with the help of a beam of protons directed on a Lithium ($^7{\rm Li}$) target. The proton beam was tuned in such a way that  
the different $^8{\rm Be}$ excitations could be separated with high accuracy.

In the data collection stage, a clear anomaly was observed in the decay of $^8{{\rm Be}^*}$ with
spin-parity $J^P=1^+$ into the ground state $^8{\rm Be}$ with spin-parity $0^+$ (both with isospin $T=0$),
where $^8{{\rm Be}^*}$ had  an excitation energy of 18.15~MeV. Upon analysis of the electron-positron properties, 
the spectra of both their opening
angle $\theta$ and  invariant mass $M$ presented the characteristics of  an excess consistent with an intermediate boson $X$
being produced on-shell
 in the decay of  the $^8{{\rm Be}^*}$ state, with the $X$ object subsequently decaying into $e^+e^-$ pairs.
The best fit to the mass $M_X$ of $X$ was given as \cite{Krasznahorkay:2015iga}
$
M_X = 16.7 \pm 0.35\ \text{(stat)}\ \pm 0.5\ \text{(sys)\ MeV},
$
in correspondence of a ratio of Branching Ratios (BRs) obtained as 
$$\frac{{\rm BR}(^8{{\rm Be}^*} \to X + {^8{\rm Be}})}{{\rm BR}(^8{{\rm Be}^*} \to \gamma + {^8{\rm Be}})} \times {\rm BR}(X\to e^+ e^-) 
= 5.8 \times 10^{-6}.$$
This combination yields a statistical significance of the excess of about $6.8\,\sigma$
 \cite{Krasznahorkay:2015iga}.

An explanation of the $X$ nature was attempted by 
\cite{Feng:2016jff,Feng:2016ysn}, in the form of models featuring a new vector boson $Z'$ with a mass
$M_{Z'}$ of about 17~MeV, with vector-like couplings to quarks and leptons.
Constraints on such a new state, notably from searches for $\pi^0\to Z'+\gamma$
by the NA48/2 experiment \cite{Batley:2015lha}, require the couplings
of the $Z'$ to up and down quarks to be `protophobic', i.e., that the charges
$e {\epsilon}_u$ and $e {\epsilon}_d$ of up and down quarks -- written as multiples of the
positron charge $e$ -- satisfy the relation $2{\epsilon}_u+{\epsilon}_d \lsim 10^{-3}$
\cite{Feng:2016jff,Feng:2016ysn}. Subsequently, further studies of such models
have been performed in \cite{Gu:2016ege,Chen:2016dhm,Liang:2016ffe,Jia:2016uxs,Kitahara:2016zyb,Chen:2016tdz,Seto:2016pks,
Neves:2016ugb,Chiang:2016cyf}\footnote{An alternative explanation was given in \cite{Ellwanger:2016wfe}, wherein the $X$ was identified with a light pseudoscalar state with couplings  to up and down type quarks  about 0.3 times those of the Standard Model (SM) Higgs boson.}.

In the footsteps of this literature, we consider here an extension of the SM described by a generic $U(1)'$ group with a light gauge boson \cite{Fayet:1980rr,Fayet:1980ad,Fayet:1990wx,Fayet:2007ua,Fayet:2008cn,Fayet:2016nyc}. 
Due to the presence of two such Abelian symmetries, $U(1)_{Y} \times U(1)'$, the most general kinetic Lagrangian of the corresponding fields, $\hat B_\mu$ and $\hat B'_\mu$, respectively, allows for a gauge invariant mixing of the two field-strengths
\bea
\mathcal L_\textrm{kin} = - \frac{1}{4} \hat F_{\mu\nu} \hat F^{\mu\nu} - \frac{1}{4} \hat F'_{\mu\nu} \hat F^{'\mu\nu} - \frac{\kappa}{2} \hat F'_{\mu\nu} \hat F^{\mu\nu},
\eea
where $\kappa$ is the kinetic mixing parameter between $U(1)_Y$ and $U(1)'$. A diagonal form for this Lagrangian can be obtained by transformation of  the Abelian fields such that the  gauge covariant derivative becomes
\be 
{\cal D}_\mu = \partial_\mu + .... + i g_1 Y B_\mu + i (\tilde{g} Y + g' z) B'_\mu , 
\ee
where $Y$ and $z$ are the hypercharge and $U(1)'$ charge, respectively, and $\tilde{g}$ the gauge coupling mixing between the two Abelian groups. 

We also consider the presence of two $SU(2)$ (pseudo)scalar doublets, embedded in a 2-Higgs Doublet Model (2HDM) scalar potential,  $\Phi_1$ and $\Phi_2$,
with the same hypercharge $Y=1/2$ and two different charges $z_{\Phi_1}$ and $z_{\Phi_2}$ under the extra $U(1)'$. 
The new Abelian symmetry replaces the discrete $Z_2$ one usually imposed in 2HDMs to avoid tree-level flavour changing neutral currents \cite{Ko:2012hd,Ko:2013zsa}.
Alongside spontaneous Electro-Weak Symmetry Breaking
(EWSB) of the SM gauge symmetry through the Vacuum Expectation Values 
(VEVs)  of the two Higgs doublets $\langle \Phi_{1,2} \rangle = v_{1,2}$, with $v^2 = v_1^2 + v_2^2$ and $\tan \beta = v_2/v_1$, 
one may in principle also have a contribution to $U(1)'$ symmetry breaking through the VEV $\langle \chi \rangle = v'$ of an extra SM-singlet scalar $\chi$, indeed connected to the mass term $m_{B'} = g' z_\chi v'$.  
The diagonalisation of the mass matrix of neutral gauge bosons implies the following mixing angle, $\theta'$, between the SM $Z$ and new $Z'$:
\bea
\label{eq:mixing2Higgs}
\tan 2 \theta' =  \frac{2  \, g_\Phi \, g_Z}{ g_{\Phi^2} + 4 m_{B'}^2/v^2 - g_Z^2},
\eea
where $g_Z = \sqrt{g_1^2 + g_2^2}$ is the EW coupling. 
The parameters $g_\Phi$ and $g_{\Phi^2}$ are defined as  
\bea
g_\Phi &=& (\tilde g + 2 g' z_{\Phi_1}) \cos^2 \beta + (\tilde g + 2 g' z_{\Phi_2}) \sin^2 \beta \nn \\
g_{\Phi^2} &=& (\tilde g + 2 g' z_{\Phi_1})^2 \cos^2 \beta + (\tilde g + 2 g' z_{\Phi_2})^2 \sin^2 \beta \,.
\eea
The $Z-Z'$ mixing $\theta'$ is generated by both the kinetic mixing $\tilde g$ and the mass mixing induced by the $Z'$ gauge interactions with the two Higgs doublets.
The general expressions of the masses of the $Z$ and $Z'$ gauge bosons are
\bea
\label{eq:ZZpmass1Higgs}
M_{Z,Z'} = g_Z \frac{v}{2} \left[ \frac{1}{2} \left( \frac{g_{\Phi^2} + 4 m_{B'}^2/v^2 }{g_Z^2} + 1\right) \mp \frac{ g_{\Phi}}{\sin 2 \theta' \, g_Z} \right]^\frac{1}{2}
\eea
and, for $g', \tilde g \ll 1$ and $m_{B'}^2 \ll v^2$, the $Z'$ mass is given by
\bea
\label{eq:Zpmass}
M_{Z'}^2 \simeq   m_{B'}^2 +  \frac{v^2}{4} {g'}^2 (z_{\Phi_1} - z_{\Phi_2})^2 \sin^2 (2 \beta),
\eea
which is non-vanishing even when $m_{B'} \rightarrow 0$ due to a possible split between the $U(1)'$ charges of the scalar doublets $z_{\Phi_1}$ and $z_{\Phi_2}$. 
In the $m_{B'} \simeq 0$ limit (equivalent to $v' \simeq 0$), in which there is no contribution from the SM-singlet $\chi$, one finds, for $M_{Z'} \simeq 17$ MeV and $v \simeq 246$ GeV, $g' \sim 10^{-4}$. 
Here, two comments are in order. Firstly, in case of one Higgs doublet (which is obtained from Eq. (\ref{eq:Zpmass}) by neglecting the second term), the limit  $m_{B'} \ll v$ leads to $M_Z' \simeq m_{B'}$ and the SM Higgs sector does not play any role in the generation of the $Z'$ mass. 
Secondly, in the 2HDM case with $z_{\Phi_1} \neq z_{\Phi_2}$, the symmetry breaking of the $U(1)'$ can actually be realised without the extra SM-singlet $\chi$, namely with $v' = 0$. In this paper we focus on the latter scenario in which the contribution of the SM-singlet $\chi$ can be completely neglected.
In this case, the typical CP-odd state of the 2HDM extensions represents the longitudinal degree of freedom of the $Z'$.

Having discussed the spontaneous symmetry breaking of the $U(1)'$, we now briefly comment on the fermion sector and the constraints imposed by the new gauge symmetry.
The conditions required by the cancellation of gauge and gravitational anomalies, which strongly constrain the charge assignment of the SM spectrum under the extra $U(1)'$ gauge symmetry, are here imposed. 
This implies the introduction of SM-singlet fermions, $s_i$, which could be exploited, in some scenarios, for implementing a seesaw mechanism generating light neutrino masses. The actual charges and masses of these new states, if present, strongly depend on the specific realisation of the fermion sector. These SM-singlet states are irrelevant for the explanation of the Atomki anomaly and will not be addressed in this paper.
The general charge assignments of the spectrum in our extension of the SM are given in Tab.~\ref{tab:charges}, where the $z_{s_i}$'s are chosen to cancel the anomaly in the $U(1)'U(1)'U(1)'$ and $U(1)'GG$ triangle diagrams, $G$ being the gravitational current. 

\begin{table}[!t]
\vspace{0.5cm}
\centering
\begin{tabular}{|c|c|c|c|c|}
\hline
		& $SU(3)$ & $SU(2)$ & $U(1)_Y$ & $U(1)'$ \\ \hline
$Q_L$	&  3		&	2	& 1/6		&	$z_Q$ \\
$u_R$	&  3		&	1	& 2/3		&	$z_u$ \\
$d_R$	&  3		&	1	& -1/3	&	$2 z_Q - z_u$\\
$L$		&  1		&	2	& -1/2	&	$-3 z_Q$ \\
$e_R$	&  1		&	1	& -1		&	$-2 z_Q - z_u$ \\
$s_i$	&  1		&	1	& 0		&	$z_{s_i}$ \\
\hline
\end{tabular}
\caption{Flavour universal charge assignment in the $U(1)'$ extension of the SM. \label{tab:charges}}
\end{table}

The interactions between the SM fermions and the $Z'$ gauge boson are described by the corresponding Lagrangian, 
$\mathcal L_\textrm{\rm int} = - J^\mu_{Z'} Z'_\mu$, where the gauge current is given by
\bea
J^\mu_{Z'} = \sum_f \bar \psi_f \gamma^\mu \left( C_{f, L} P_L + C_{f, R} P_R \right) \psi_f
\eea
with left- and right-handed coefficients
\bea
C_{f,L} &=&  - g_Z s' \left( T^3_f - s_W^2 Q_f \right) + ( \tilde g Y_{f, L} + g' z_{f, L})  c' \,, \nn \\
C_{f,R} &=&  g_Z s_W^2 s' Q_f + ( \tilde g Y_{f, R} + g' z_{f, R}) \, c' \,.
\eea
In these equations we have adopted the shorthand notations $s_W \equiv \sin \theta_W$, $c_W \equiv \cos \theta_W$, $s' \equiv \sin \theta'$ and $c' \equiv \cos \theta'$. 
We have also introduced $Y_f$ the hypercharge, $z_f$ the $U(1)'$ charge, $T^3_f$ the third component of the 
weak isospin and $Q_f$ the electric charge of a generic fermion $f$.
Analogously, the vector and axial-vector components of the $Z'$ interactions are
\begin{widetext}
\bea
C_{f, V} &=& \frac{C_{f, R} + C_{f, L}}{2} = \frac{1}{2} \left[ 
- g_Z s' (T^3_f - 2 s_W^2 Q_f) + c' \tilde g (2 Q_f - T^3_f) + c' g' (z_{f, L} + z_{f,R})
\right] \,, \nn \\
C_{f, A} &=& \frac{C_{f, R} - C_{f, L}}{2} = \frac{1}{2} \left[ 
 (g_Z s'  + \tilde g c' ) T^3_f - c' g' (z_{f, L} - z_{f, R}) 
\right] \,,
\eea
\end{widetext}
where we have exploited the relation $Y_f = Q_f - T^3_f$. 
These equations can considerably be simplified by realising that $g_Z s'$ is of the same order of $\tilde g$ for $g', \tilde g \ll 1$, which leads to
\bea
\label{eq:CVA_expanded}
C_{f, V} &\simeq&    \tilde g  c_W^2 \, Q_f + g'  \left[ z_\Phi (T^3_f - 2 s_W^2 Q_f)  + z_{f,V} \right] \,, \nn \\
C_{f, A} &\simeq&  g' \left[   -  z_\Phi \, T^3_f  +   z_{f,A} \right],
\eea
where we have introduced the vector and axial-vector $U(1)'$ charges $z_{f,V/A} = 1/2(z_{f,R} \pm z_{f,A})$.
Notice that $z_\Phi$ can be either $z_H$ for a single Higgs doublet ($H$) model or a combination of $z_{\Phi_1}$ and $z_{\Phi_2}$, namely $z_\Phi = z_{\Phi_1} \cos^2 \beta +  z_{\Phi_2} \sin^2 \beta$, for a 2HDM. 
 The $z_\Phi$ charge arises from the small gauge coupling expansion of the $Z-Z'$ mixing angle $\theta'$ which implies $c' \simeq 1$ and $s' \simeq - g_\Phi/g_Z = - \tilde g - 2 g' z_\Phi$.
The $Z'$ couplings are characterised by the sum of three different contributions. 
The kinetic mixing $\tilde g$ induces a vector-like term proportional to the Electro-Magnetic (EM) current which is the only source of interactions when all the SM fields are neutral under $U(1)'$. In this  case the $Z'$ is commonly dubbed \emph{dark photon}. The second term is induced by  $z_\Phi$, the $U(1)'$ charge in the Higgs sector, and leads to a \emph{dark Z}, namely a gauge boson mixing with the SM $Z$ boson. Finally, there is the standard gauge interaction proportional to the fermionic $U(1)'$ charges $z_{f,V/A}$. 

We can now delineate two different scenarios depending on the structure of the axial-vector couplings of the $Z'$ boson. 
In particular, when only a $SU(2)$ doublet is taken into account, the $C_{f,A}$ coefficients are suppressed with respect to the vector-like counterparts (see  \cite{Kahn:2016vjr}). 
This is a direct consequence of the gauge invariance of the Yukawa interactions which forces the $U(1)'$ charge of the Higgs field to satisfy the conditions $z_H = z_{Q} - z_{d} = - z_{Q} + z_{u} = z_L - z_e$.
Inserting the previous relations into Eq. (\ref{eq:CVA_expanded}), we find $C_{f, A} \simeq 0$, which describes a $Z'$ with only vector interactions with charged leptons and quarks.
We stress again that the suppression of the axial-vector coupling is only due to the structure of the scalar sector, which envisions only one $SU(2)$ doublet, and to the gauge invariance of the Yukawa Lagrangian. 
This feature is completely unrelated to the $U(1)'$ charge assignment of the fermions, the requirement of anomaly cancellation and  the matter content potentially needed to account for it. 
In the scenario characterised by two Higgs doublets, the axial-vector couplings of the $Z'$ are, in general, of the same order of magnitude as the vector ones and the cancellation between the two terms of $C_{f, A}$ in Eq.(\ref{eq:CVA_expanded}) is not achieved regardless of the details of the Yukawa Lagrangian, namely, of the type of the 2HDM considered. 
Notice that, unlike the pure vector-like case, it is extremely intricate to build up a model of a $Z'$ with only axial-vector interactions and, in general, both $C_{V}$ and $C_{A}$ are present.

A well-known realisation of the scalar sector with two Higgs doublets is the so-called type-II in which the up-type quarks couple to one Higgs doublet (conventionally chosen to be $\Phi_2$) while the down-type quarks couple to the other ($\Phi_1$). 
The anomaly cancellation condition arising from the $U(1)'SU(3)SU(3)$ diagram and the gauge invariance of the Yukawa Lagrangian require $2 z_Q - z_d - z_u = z_{\Phi_1} - z_{\Phi_2} = 0$, which
necessarily calls for extra coloured states when $z_{\Phi_1} \neq z_{\Phi_2}$. These new states must be vector-like under the SM gauge group and chiral under the extra $U(1)'$ \cite{Kahn:2016vjr}.

\begin{figure}
\includegraphics[width=7.cm,height=7.cm]{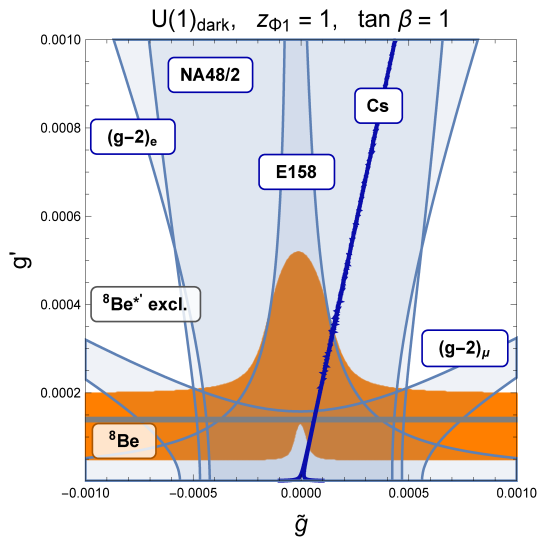}
\caption{
Allowed parameter space (orange region) explaining the anomalous $\Be^*$ decay. 
The white region above is excluded by the non-observation of the same anomaly in the ${\Be^{*}}'$ transition. Also shown (shaded regions) is the allowed parameter space by the $g-2$ of electrons and muons and the M{\o}ller scattering at SLAC E158. The blue line selects values of $g'$ and $\tilde g$ compatible with the weak nuclear charge measurement of Cesium. The horizontal grey band delineates values of $g'$ for which the $Z'$ mass is solely generated by the SM vev.
\label{fig:typeI}}
\end{figure}

\begin{figure}[ht]
\begin{center}
\includegraphics[width=4.cm,height=2.cm]{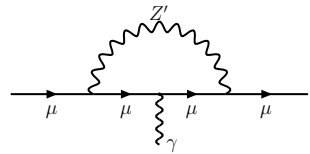}~~~~~
\caption{The new contribution to the muon anomalous magnetic
moment in a $U(1)'$ extension of the SM.} \label{g-2}
\end{center}
\end{figure}

A far more interesting scenario is realised when the scalar sector reproduces the structure of the type-I 2HDM in which only one ($\Phi_2$) of the two Higgs doublets participates in the Yukawa interactions. The corresponding Lagrangian is the same as the SM one and its gauge invariance simply requires $z_{\Phi_2} = - z_Q + z_u = z_{Q} - z_d = z_{L} - z_e$, without constraining the $U(1)'$ charge of $\Phi_1$.
Differently from the type-II scenario in which extra coloured states are required to build an anomaly-free model, in the type-I case the UV consistency of the theory can be easily satisfied introducing only SM-singlet fermions as demanded by the anomaly cancellation conditions of the $U(1)'U(1)'U(1)'$ and $U(1)'GG$ correlators. Nevertheless, the mismatch between $z_\Phi$ and $z_{f,A}=\pm z_{\Phi_2}/2$ (for up-type and down-type quarks, respectively) prevents $C_{f,A}$ to be suppressed 
and the $Z'$ interactions are given by
\begin{align}
&C_{u, V} =  \frac{2}{3} \tilde g  c_W^2  + g'  \left[ z_\Phi \left(\frac{1}{2} -  \frac{4}{3}  s_W^2 \right)  + z_{u,V} \right], \nn\\
&C_{u, A} = - \frac{g'}{2} \cos^2 \beta (z_{\Phi_1} - z_{\Phi_2})  \,, \nn \\
&C_{d, V} =  -\frac{1}{3} \tilde g  c_W^2  + g'  \left[ z_\Phi \left(-\frac{1}{2} +  \frac{2}{3} s_W^2  \right)  + z_{d,V} \right],  \nn\\
&C_{d, A} = \frac{g'}{2} \cos^2 \beta  (z_{\Phi_1} - z_{\Phi_2}) \,, \nn \\
&C_{e, V} =  - \tilde g  c_W^2  + g'  \left[ z_\Phi \left(-\frac{1}{2} +  2 s_W^2  \right)  + z_{e,V} \right],  \nn\\
&C_{e, A} =  \frac{g'}{2} \cos^2 \beta  (z_{\Phi_1} - z_{\Phi_2}) \,, \nn \\
&C_{\nu, V} = - C_{\nu, A} = \frac{g'}{2} \left( z_{\Phi} + z_L \right) .
\label{couplings}
\end{align}
We now show that the structure of the couplings discussed above is able to explain the presence of a $Z'$ resonance in the $\Be^*$ decay.
As pointed out in \cite{Feng:2016ysn}, the contribution of the axial-vector couplings to the $\Be^* \rightarrow \Be \, Z'$ decay is proportional to $k/M_{Z'} \ll 1$, where $k$ is the momentum of the $Z'$, while the vector component is suppressed by $k^3/M_{Z'}^3$. Therefore, in our case, being $C_{f,V} \sim C_{f,A}$, we can neglect the effects of the vector couplings of the $Z'$ and their interference with the axial counterparts. The relevant matrix elements for the $\Be^*$ transition mediated by an axial-vector boson have been computed in \cite{Kozaczuk:2016nma}.
Notice that the axial couplings of the quarks and, therefore, the width of the $\Be^* \rightarrow \Be \, Z'$ decay are solely controlled by the product $g' \cos^2 \beta$ while the kinetic mixing $\tilde g$ only affects the $\textrm{BR}(Z' \rightarrow e^+e^-)$ since the $Z' \rightarrow \nu \nu$ decay modes are allowed (we assume that the $Z' \rightarrow s_i s_j$ decays are kinematically closed).
For definiteness, we consider a $U(1)_\textrm{dark}$ charge assignment with $z_{f} =0$ and $z_{\Phi_2} = 0$ and we choose $z_{\Phi_2} = 1$ and $\tan \beta = 1$.
Analogue results may be obtained for different $U(1)'$ charge assignments and values of $\tan \beta$.
We show in Fig.~\ref{fig:typeI} the parameter space explaining the Atomki anomaly together with the most constraining experimental results.
The orange region, where the $Z'$ gauge couplings comply with the best-fit of the $\Be^*$ decay rate in the mass range $M_{Z'} = 16.7 \, {\rm MeV} - 17.6 \, {\rm MeV}$ \cite{Krasznahorkay:2015iga,Feng:2016ysn}, encompasses the uncertainties on the computation of the nuclear matrix elements \cite{Kozaczuk:2016nma}. The 
region above it is excluded by the non-observation of the same transition in the isovector excitation ${\Be^{*}}'$ \cite{Krasznahorkay:2015iga} 
\footnote{The Atomki collaboration has recently presented evidences of an excess, compatible with a 17 MeV boson mediation, in the ${\Be^{*}}'$ transition at the ``International Symposium: Advances in Dark Matter and Particle Physics 2016" \cite{Krasznahorkay:2017gwn}. Since this result is not public we do not account for it, but we ought to mention it.}. 
The horizontal grey band selects the values of $g'$ accounting for the $Z'$ mass in the negligible $m_{B'}$ case in which the $U(1)'$ symmetry breaking is driven by the two Higgs doublets.
Furthermore, among all other experimental constraints involving a light $Z'$ that may be relevant for this analysis we have shown the most restrictive ones. The parity-violating M{\o}ller scattering measured at the SLAC E158 experiment \cite{Anthony:2005pm} imposes a constraint on the product $C_{e,V} C_{e,A}$ of the $Z'$, namely $|C_{e,V} C_{e,A}| \lesssim 10^{-8}$ for $M_{Z'} \simeq 17$ MeV \cite{Kahn:2016vjr}. 
The strongest bound comes from the atomic parity violation in Cesium (Cs), namely from the measurement of  its weak nuclear charge $\Delta Q_W$ \cite{Davoudiasl:2012ag,Bouchiat:2004sp},
which requires $| \Delta Q_W | \lesssim 0.71$ at $2\sigma$ \cite{Porsev:2009pr}.
It represents a constraint on the product of $C_{e,A}$ and a combination of $C_{u,V}$ and $C_{d,V}$.
{
This bound can be avoided if the $Z'$ has either only vector or axial-vector couplings but in the general scenario considered here, it imposes severe constraints on the gauge couplings $g',\tilde g$ thus introducing a fine-tuning in the two gauge parameters.
}
{
Also shown is the constraint from neutral pion decay, $\pi \rightarrow Z' \gamma$, from the NA48/2 experiment \cite{Raggi:2015noa}. The bound is proportional to the anomaly factor $|C_{u, V} Q_u- C_{d, V} Q_d|^2$ which depends on the vector coupling of the $Z'$ and is given by $|2 C_{u, V} + C_{d, V} | \lesssim 3.6 \times 10^{-4}/\sqrt{\textrm{BR}(Z' \rightarrow e^+e^-)}$ for $M_{Z'} \simeq 17$ MeV. The contribution of the axial components is induced by chiral symmetry breaking effects and is, therefore, suppressed by the light quark masses.
}
%
%
%
The light-boson contribution to the anomalous magnetic moment of the electron has also been taken into account, as it is required to be within the $2\sigma$ uncertainty of the departure of the SM prediction from the experimental result \cite{Giudice:2012ms}.
We now analyse the contribution of a very light gauge boson $Z'$ to the muon
anomalous magnetic moment \cite{Altmannshofer:2016brv} which has been measured at
 Brookhaven National Laboratory to a precision of $0.54$ parts per million. The current
average of the experimental results is given by \cite{Bennett:2006fi,Blum:2013xva,Lindner:2016bgg} 
\be%
a^{\rm exp}_{\mu}=11 659 208.9(6.3)\times 10^{-10},%
\ee%
which is different from the SM prediction by $3.3
\sigma$ to $3.6\sigma$: 
$
\Delta a_{\mu}=a^{\rm exp}_{\mu}-a^{\rm SM}_{\mu}=(28.3 \pm
8.7\ {\rm to}\ 28.7 \pm
8.0)\times 10^{-10}.
$
From the interaction Lagrangian described above
one finds a new contribution to  $(g-2)_\mu$ generated by a one-loop diagram with $Z'$ exchange as shown in Fig.~\ref{g-2}, which leads to
%
\be%
\delta a^{Z'}_{\mu}= \frac{r_{m_\mu}}{4\pi^2} \left[ C_{\mu,V}^2  \, g_V(r_{m_\mu}) - C_{\mu,A}^2  \, g_A(r_{m_\mu}) \right],
\ee%
where $r_{m_\mu} \equiv (m_\mu/M_{Z'})^2$ and $g_V,g_A$ are given by%
\begin{align}
& g_V(r)= \int^1_0 dz\frac{z^2(1-z)}{1-z+r z^2}, \\
& g_A(r)= \int^1_0 dz\frac{(z-z^2)(4-z) + 2r z^3}{1-z+r z^2} \,.
\end{align}
For $M_{Z'} \simeq 17$ MeV one finds $\delta a^{Z'}_{\mu} \simeq 0.009 \, C_{\mu, V}^2 - \, C_{\mu, A}^2$.
 We require again that the contribution of the $Z'$ to  $(g-2)_\mu$, which is mainly due to its axial-vector component, is less than the $2\sigma$ uncertainty of the discrepancy between the SM result and the experimental measure. 

We finally comment on the constraints imposed by neutrino-electron scattering processes \cite{Vilain:1994qy,Deniz:2009mu,Bellini:2011rx}, the strongest one being from $\bar \nu_e e$ scattering at the TEXONO experiment \cite{Deniz:2009mu}, which affect a combination of $C_{e, V/A}$ and $C_{\nu,V}$. 
In the protophobic scenario, in which the $Z'$ has only vector interactions, the constrained $\nu$ coupling to the $Z'$ boson is in high tension with the measured $\Be^*$ decay rate since $C_{\nu,V} = -2 C_{n,V}$, where $C_{n,V} = C_{u,V} + 2 C_{d,V}$ is the coupling to neutrons, and a mechanism to suppress the neutrino coupling must be envisaged \cite{Feng:2016ysn}. 
This bound is, in general, alleviated if the one attempts to explain the Atomki anomaly with a $Z'$ boson with axial-vector interactions since the required gauge couplings $g',\tilde g$ are smaller than the ones needed in the protophobic case. 
{
Neutrino couplings are also constrained by meson decays, like, for instance $K^\pm \rightarrow \pi^\pm \nu \nu$ which has been studied in \cite{Davoudiasl:2014kua} and where it has been shown that the corresponding constraint is relaxed by a destructive interference effect induced by the charged Higgs. As the results presented in \cite{Davoudiasl:2014kua} relies on the Goldstone boson equivalence approximation,
we have computed the full one-loop corrections to the $K^\pm \rightarrow \pi^\pm Z'$ process in the $U(1)'$-2HDM scenario. The details will be presented in a forthcoming work and the results are in agreement with the estimates in \cite{Davoudiasl:2014kua}.
In our setup, for $g' \sim 10^{-4}$ and $\tan \beta = 1$, $M_{H^\pm} \sim 600$ GeV can account for the destructive interference quoted above between the $W^\pm$ and $H^\pm$ loops. 
For instance, we find $\textrm{BR}(K^\pm \rightarrow \pi^\pm Z' \rightarrow \pi^\pm \nu \nu) \simeq 0.1 \, \textrm{BR}(K^\pm \rightarrow  \pi^\pm \nu \nu)_\textrm{exp}$ for $M_{H^\pm} \sim 615$ GeV with $\textrm{BR}(Z' \rightarrow \nu\nu) \simeq 30\%$ which is the maximum value for the invisible $Z'$ decay rate in the allowed region (orange and grey shaded area) shown in Fig.~\ref{fig:typeI}. 
A similar constraint arises from the $B$ meson decay to invisible but is less severe than the one discussed above \cite{Patrignani:2016xqp}.
The $B^\pm \rightarrow K^\pm Z'$ process is characterised by the same loop corrections appearing in $K^\pm \rightarrow \pi^\pm Z'$, with the main difference being the dependence on the CKM matrix elements. Therefore, the suppression effect induced by the charge Higgs mass affects both processes in the same region of the parameter space, thus ensuring that the bound from the invisible $B$ decays is satisfied once the constraint from the analogous $K$ meson decay is taken into account.
} \\
In summary, we have come to an exciting conclusion. The model that we have constructed, which minimally departs from the SM,  in both the gauge sector (wherein a dark $U(1)'$ is added) and Higgs framework (wherein a second doublet is added with a type-I Yukawa configuration), with the two intertwined as it is the pseudoscalar state of the latter that spontaneously breaks the symmetry of the former, has the potential to explain the 
anomaly in the decays of the Beryllium. Notably, the ballpark of values of the $g'$ coupling reproducing the $\Be$ internal pair creation excess also predicts the mass of the $Z'$ from EWSB and, therefore, as for the masses of the $Z$ and $W$ gauge bosons, the $M_{Z'} = 17$ MeV could be generated by the same EW mass scale $v \simeq 246$ GeV.

\section*{Acknowledgements}
The work of LDR and SM is supported in part by the NExT Institute. SM also acknowledges partial financial contributions from the STFC Consolidated Grant ST/L000296/1. Furthermore, the work of LDR has been supported by the STFC/COFUND Rutherford International Fellowship scheme. 
SK was partially supported by the STDF project 13858 and the European Union's Horizon 2020 research and innovation programme under the Marie Sklodowska-Curie grant agreement No. 690575.
All authors finally acknowledge support from the grant H2020-MSCA-RISE-2014 No. 645722 (NonMinimalHiggs).

\end{document}